\documentclass{article}
\usepackage{main}

\usepackage{times}
\usepackage{soul}
\usepackage{url}
\usepackage[hidelinks]{hyperref}
\usepackage[utf8]{inputenc}
\usepackage[small]{caption}
\usepackage{graphicx}
\usepackage{amsmath}
\usepackage{amsthm}
\usepackage{booktabs}
\usepackage{algorithm}
\usepackage{algpseudocode}
\usepackage{multirow}
\usepackage[switch]{lineno}
\usepackage{color}

\title{FG-RAG: Enhancing Query-Focused Summarization with Context-Aware Fine-Grained Graph RAG}

\author{
Yubin Hong$^1$
\and
Chaofan Li$^1$\and
Jingyi Zhang$^1$\And
Yingxia Shao$^1$\\
\affiliations
$^1$Beijing University of Posts and Telecommunications\\
\emails
bupt.hongyubin@gmail.com,
cfli@bupt.edu.cn,
bupt.zhangjingyi@gmail.com,
shaoyx@bupt.edu.cn
}

\begin{document}

\maketitle

\begin{abstract}
    Retrieval-Augmented Generation (RAG) enables large language models to provide more precise and pertinent responses by incorporating external knowledge. In the Query-Focused Summarization (QFS) task, GraphRAG-based approaches have notably enhanced the comprehensiveness and diversity of generated responses. However, existing GraphRAG-based approaches predominantly focus on coarse-grained information summarization without being aware of the specific query, and the retrieved content lacks sufficient contextual information to generate comprehensive responses. To address the deficiencies of current RAG systems, we propose Context-Aware Fine-Grained Graph RAG (FG-RAG) to enhance the performance of the QFS task. FG-RAG employs Context-Aware Entity Expansion in graph retrieval to expand the coverage of retrieved entities in the graph, thus providing enough contextual information for the retrieved content. Furthermore, FG-RAG utilizes Query-Level Fine-Grained Summarization to incorporate fine-grained details during response generation, enhancing query awareness for the generated summarization. Our evaluation demonstrates that FG-RAG outperforms other RAG systems in multiple metrics of comprehensiveness, diversity, and empowerment when handling the QFS task. Our implementation is available at \href{https://github.com/BuptWululu/FG-RAG}{https://github.com/BuptWululu/FG-RAG}.
\end{abstract}

\section{Introduction}

Retrieval-Augmented Generation (RAG) systems enhance the capabilities of large language models (LLMs) by integrating external knowledge bases~\cite{rag,rag0,rag2,rag1}. The integration allows LLMs to access up-to-date information~\cite{time1,time2,time3}, thereby reducing the risk of generating inaccurate or ``hallucinated" responses by using retrieved data as contexts~\cite{hallucination1,hallucination2}. RAG systems improve the precision, currency, and clarity of the responses generated by LLMs.

Query-Focused Summarization (QFS) task~\cite{qfs} aims to create summaries from documents by extracting or generating content relevant to a specific query. It involves identifying relevant entities from extensive documents and understanding and synthesizing the intricate relationships between them~\cite{qfs1,qfs2,qfs3}.
Consider the query, ``How can beekeepers market and sell their honey and other hive products?” We need to sift through the mass of documents to find the contents that discuss how to bring bee products to market, such as utilizing social media, attending farmers' markets, creating attractive packaging, or setting up an online store. It is challenging to accurately discern which pieces of information are relevant to marketing and selling activities and synthesize them into a coherent, concise summary that answers the query effectively.

In order to enhance the capacity of RAG systems to handle the QFS task, an effective approach is to transform the external knowledge into a graph~\cite{kg1,kg2}, retrieve relevant contents for the query from the graph, and synthesize them into a coherent response using LLMs, such as GraphRAG~\cite{graphrag} and LightRAG~\cite{lightrag}. These methods simplify the analysis of complex relationships among query entities in the QFS task through the use of a graph structure, thus optimize the comprehensiveness and diversity of the responses. However, these GraphRAG-based approaches suffer from the following problems:

1) \textbf{Insufficient contexts for the graph retrieval.} The QFS task requires the RAG system to perform extensive searches in the knowledge base to cover multiple perspectives and levels of information, thus ensuring that the generated answers are comprehensive and in-depth. We observe that the LLM has limited understanding of the domain while generating domain-specific responses~\cite{domain}. It is challenging to produce coherent and organized answers to queries within the QFS task if the retrieved content from the graph does not contain enough contextual information. For example, when LightRAG retrieves relevant content for the previously mentioned query, it successfully finds a considerable amount of information directly related to honey, beekeepers, etc. However, this content lacks background information on the local history and geographical context of the honey's place of origin, which would enable the LLM to address queries with a broader array of perspectives and insights.

2) \textbf{Unawareness of the query in summarization.} Existing GraphRAG-based approaches predominantly focus on coarse-grained information summarization, which is unaware of the specific query. For GraphRAG, the summarization of entities is a synthesis of the entities' intrinsic information, but not all the information contributes meaningfully to a certain query, as it may introduce unnecessary noise. For instance, in the preceding example, GraphRAG might retrieve information about the nutritional content of honey, its processing, and so forth; however, this data is irrelevant to resolving the query. 
Such information, which is not useful for responding to queries, not only affects the quality of LLM generation, but also inevitably increases token overheads.
Truly essential for query responses is the fine-grained information that is highly relevant to the query, highlighting a fundamental mismatch between generated summarization and query response requirements.

In this paper, we introduce Context-Aware Fine-Grained Graph RAG (FG-RAG) to enhance the performance of the QFS task. 
FG-RAG utilizes Context-Aware Entity Expansion during graph retrieval, employing retrieved entities to perform a complementary search for discovering additional related entities and exploring the relationships among them. Through this process, we expect to increase the coverage of retrieved entities in the graph, thus providing more contextual information for the retrieved content and improving the comprehensiveness of the subsequently generated summaries. 
Furthermore, Query-Level Fine-Grained Summarization utilized by FG-RAG is designed to consider the specific query, ultimately generating several highly relevant summaries of the query.
This method improves response accuracy by synthesizing the coarse-grained information in the graph into summaries that are highly relevant to the query, reducing the incorporation of noise. Our contributions are summarized as follows:

\begin{itemize}
    \item We propose incorporating Context-Aware Entity Expansion into the graph retrieval, which is a novel approach to enrich the contextual information of the retrieved content and provide more domain background for LLM generation, thus improving the quality of the output.
    \item We propose an innovative method that effectively integrates coarse-grained information in the graph into fine-grained information that is highly relevant to the query, reduces the inclusion of noise, and improves the accuracy of LLM-generated responses.
    \item Extensive experiments on the QFS task demonstrate that FG-RAG outperforms current state-of-the-art RAG methods in terms of comprehensiveness, diversity, and empowerment, while reducing token overhead, and also exhibits excellent adaptability in different tasks.
\end{itemize}

\begin{figure*}[!t]
    \centering
    \includegraphics[width=1\textwidth]{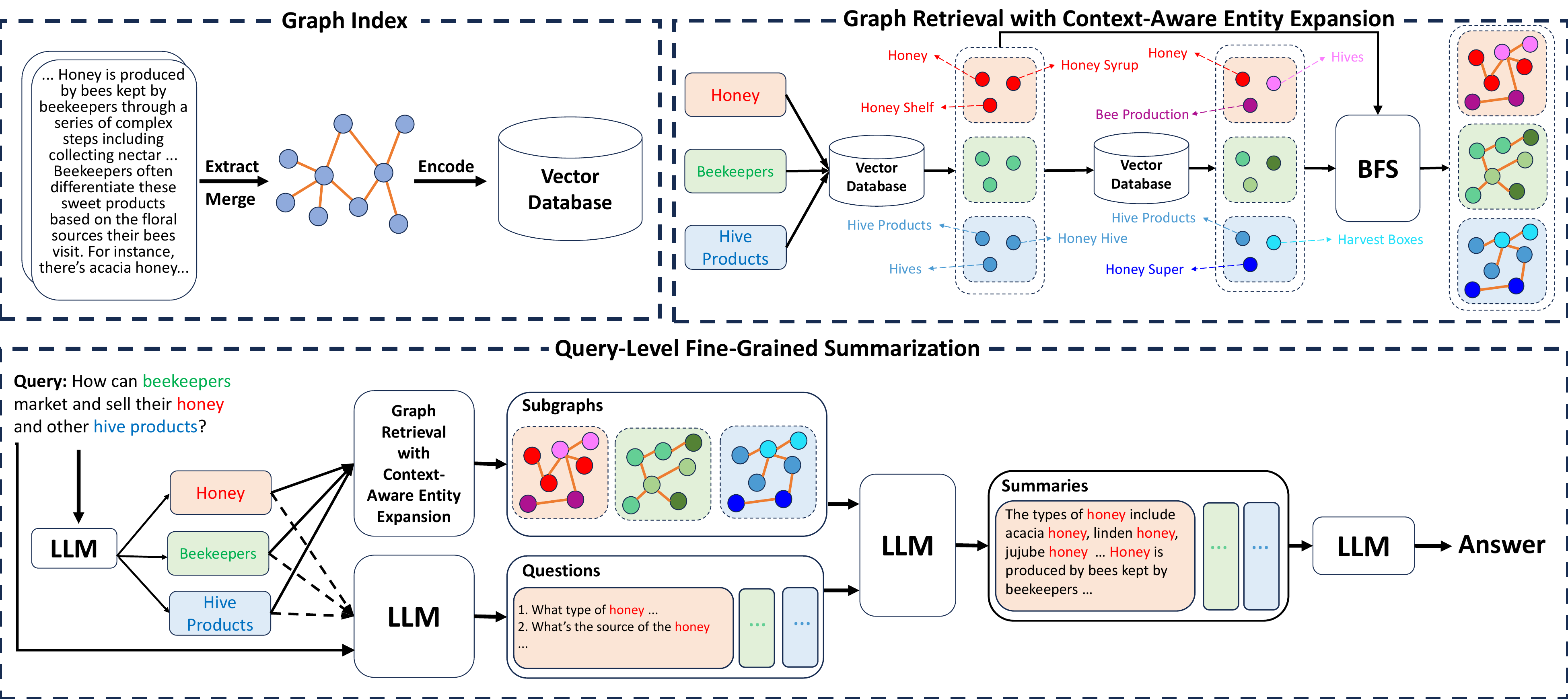}
    \caption{The overall workflow of FG-RAG. In Query-Level Fine-Grained Summarization, for each entity mentioned in the query, we sequentially perform the tasks of asking relevant questions, retrieving pertinent descriptions, and generating summaries. These summaries are then provided to the LLM, which uses them to formulate the answer.}
    \label{fig:framework}
    \vspace{-0.1in}
\end{figure*}

\section{Related Work}

\subsection{Chunk-based Retrieval-Augmented Generation}

Traditional RAG systems~\cite{traditional} improve the accuracy and specificity of LLM generation by retrieving query-relevant content from an external knowledge base and adding the retrieved content to the context window of the LLM, and easily enable rapid updating of knowledge without retraining or fine-tuning the model. Naive RAG~\cite{naiverag} typically divides the source document into chunks and embeds these chunks into vector space. For retrieving chunks relevant to a query, simply embed the query into vector space and select the first k nearest chunks. 

Many RAG systems are further enhanced based on Naive RAG, including RQ-RAG~\cite{rqrag}, Self-RAG~\cite{self}, and others~\cite{memorag,radit,sail}. RQ-RAG is an advanced RAG system designed to refine search queries dynamically. It equips the model with the ability to rewrite, decompose, and disambiguate queries. Self-RAG introduces a mechanism for self-reflection into the RAG framework. It allows the model to assess whether it needs to retrieve additional information before generating an answer, evaluate the relevance of retrieved documents, and critique the generated responses. Although these advanced RAG methods further improve the accuracy of LLM generation, their effectiveness is limited by the fact that they can only retrieve a subset of documents. This limitation hinders the comprehensive acquisition of global information, posing challenges in solving the QFS task.

\subsection{Graph-based Retrieval-Augmented Generation}
In contrast to traditional RAG, which embeds the source document into vector space following its division into chunks, graph-based RAG identifies the entities within the document along with their interrelations, and employs this information to refine query retrieval.

Different graph-based RAG methods utilize the information in the graph in different ways. GraphRAG~\cite{graphrag} builds communities from the information in the graph and determines the relevance of the community summary to the query through the LLM. LightRAG~\cite{lightrag} combines graph structure with vector representation and employs a dual-level retrieval paradigm, which greatly improves the retrieval efficiency and the quality of LLM generation. SubGraphRAG~\cite{subgraphrag} and \href{https://github.com/circlemind-ai/fast-graphrag}{FastGraphRAG} further improve the accuracy of these GraphRAG-based methods for solving the multi-hop QA task by identifying entities within queries and employing various algorithms to retrieve their associated triples or inference paths.
These methods aim either to provide a concise and precise response to the query or to generate a coarse-grained summary of the graph's information while neglecting the specifics of the query. Moreover, these methods often use the query itself or the entities within the query for graph retrieval, resulting in retrieved content that lacks relevant contextual information essential for solving the QFS task.

\section{FG-RAG Framework}

The overall workflow of FG-RAG is shown in Figure ~\ref{fig:framework}. The FG-RAG framework is divided into three parts. The Graph Index transforms external knowledge into a graph structure and encodes entity information into vectors to be stored in a vector database. The graph retrieval with Context-Enhanced Entity Expansion allows for the retrieval of entities containing more contextual information, improving the comprehensiveness of the retrieved content. The Query-Level Fine-Grained Summarization consolidates coarse-grained information from the graph into fine-grained information, generating summaries that are highly relevant to the query and ultimately utilized in the response.

\subsection{Graph Index}

For the source document $\mathcal{D}$, it is segmented into smaller and more manageable text chunks $\mathcal{D}_i$. For each segmented text chunk $\mathcal{D}_i$, the LLM identify entities and extract their relationships from the text. Since there are a large number of entities with the same name in the graph, to minimize the size of the graph and ensure the efficiency of the graph retrieval, we merge these entities. The above process is formally represented as follows:
\begin{align}
    \mathcal{V}, \mathcal{E} &= \cup_{\mathcal{D}_i\in\mathcal{D}} \text{Extract}(\mathcal{D}_i), \\
    \mathcal{V}' &= \text{Merge}(\mathcal{V}), \\
    \mathcal{G} &= (\mathcal{V}', \mathcal{E}).
\end{align}%
The Extract function, representing the extraction process executed by the LLM, creates the entity set $\mathcal{V}$ along with the relationship set $\mathcal{E}$. The Merge function consolidates entity descriptions with identical names into a unified collection $\mathcal{V}'$, thereby streamlining the existing entity set $\mathcal{V}$. This unified collection, containing aggregated entity descriptions, is then embedded into a vector database. The merged set of entities $\mathcal{V}'$ and the set of relationships $\mathcal{E}$ together form the final graph $\mathcal{G}$.

\subsection{Graph Retrieval with Context-Enhanced Entity Expansion}

Retrieval is the most important part of the RAG, as high-quality information retrieved is necessary for the LLM to generate accurate responses. Given the constraints of the inadequate contextual information in retrieved content, coupled with the LLM's limited domain-specific knowledge, it is difficult to comprehensively solve the QFS task from multiple perspectives with only the fragmented information in the graph.

In order to provide more contextual information and enhance LLM's understanding of in-domain knowledge, FG-RAG employs the graph retrieval with Context-Enhanced Entity Expansion to retrieve and utilize both weak-context entities and strong-context entities.

Given an input entity $E$ to be retrieved and a graph $\mathcal{G} = (\mathcal{V}, \mathcal{E})$, the graph retrieval with Context-Enhanced Entity Expansion seeks to identify a subgraph $\mathcal{G}' = (\mathcal{V}', \mathcal{E}')$ that is relevant to the entity $E$. We summarize this process formally as follows:
\begin{align}
    \mathcal{G}' = \text{Ret}(E, \mathcal{G}).
\end{align}%
To obtain this subgraph, a simple approach is to match the top-k entities or relationships from the vector database based on the similarity metric $\text{Sim}(E, \mathcal{E} \cup \mathcal{V}) \in \mathcal{R}$. This process is formally expressed as follows:
\begin{align}
    (\mathcal{V}', \mathcal{E}') = \text{Match}(E, \mathcal{G}).
\end{align}%
Despite the high similarity of both entities and relationships matched in the graph to the input entity $E$, they lack adequate contextual information necessary to thoroughly elucidate the complex relationships linking $E$ with other entities within the source document.
The core issue arises from the fact that the entity information in the graph structure transformed from external knowledge is fragmented. Relying solely on singling out a query or its entities to match entities in the graph provides insufficient contextual information to solve the QFS task.

To enhance contextual information about the input entity $E$ during graph retrieval, we incorporate Context-Aware Entity Expansion into this process. We match the input entity $E$ to top-n similar entities from the graph to obtain the entity set $\mathcal{E}_w$ via the vector database. After the first step is completed, we use $\mathcal{E}_w$ as the input to the vector database for further matching to obtain the entity set $\mathcal{E}_s$, which is formally formulated as follows:
\begin{align}
    \mathcal{E}_w &= \text{Match}(E, \mathcal{V}),\\
    \mathcal{E}_s &= \text{Match}(\mathcal{E}_w, \mathcal{V}).
\end{align}%
Since $\mathcal{E}_s$ contains more contextual information related to the query compared to $\mathcal{E}_w$, we refer to entities in $\mathcal{E}_w$ as weak-context entities and entities in $\mathcal{E}_s$ as strong-context entities. In order to further improve the comprehensiveness and diversity of the retrieval results, we conduct BFS separately on $\mathcal{E}_w$ and $\mathcal{E}_s$. The final retrieval results $\mathcal{G}' = (\mathcal{V}', \mathcal{E}')$ are then compiled by merging and removing duplicates from the descriptions gathered during the BFS traversals.

The graph retrieval with Context-Enhanced Entity Expansion employs retrieved entities to perform a complementary search for discovering additional related entities and providing more comprehensive contextual information for the retrieval results.

\subsection{Query-Level Fine-Grained Summarization}

Currently, the summaries obtained by the GraphRAG-based method are based on the LLM's overall understanding of the information in the graph, and such summaries are considered coarse-grained information. These methods focus more on organizing the retrieved information than on applying it in response to a query, which results in summarization that often include information that is not relevant to the specific query.

In order to reduce the inclusion of query-irrelevant information while making full use of the large amount of coarse-grained information in the graph, FG-RAG employs Query-Level Fine-Grained Summarization. It identifies key entities within the query, formulates questions based on these entities, and synthesizes a summary of the relevant information found in the graph. 
By doing so, we furnish the LLM with a fine-grained summary that is closely aligned with the query, thus bringing the generated responses closer to fulfilling the actual needs of the query. The following are the specific steps:
\begin{itemize}
    \item \textbf{Query entities extraction.} 
    For a given query $\mathcal{Q}$, FG-RAG first extracts the query entities in it. The formalization is as follows:
    \begin{align}
        \text{Decompose}(\mathcal{Q}) &= \{E_1, E_2, ... , E_n\}.
    \end{align}%
    Assuming that the query $\mathcal{Q}$ is ‘How can beekeepers market and sell their honey and other hive products?’ The set of query entities extracted by the LLM is [“Honey”, “Beekeepers”, “Hive Products”].
    \item \textbf{Relevant descriptions retrieval.} 
    For each query entity $E_i$, FG-RAG employs the Graph Retrieval with Context-Aware Entity Expansion to retrieve the subgraph associated with it. The formalization is as follows:
    \begin{align}
        \mathcal{G}'_i = \text{Ret}(E_i, \mathcal{G}).
    \end{align}%
    Then FG-RAG extracts descriptions of all entities and relationships from the subgraph $\mathcal{G}'_i$, serving as the basis for generating the subsequent summary $\mathcal{S}_i$.
    \item \textbf{Relevant questions formulation.} Breaking down a complex problem into smaller, manageable subproblems and addressing them individually can significantly enhance the precision of their resolution~\cite{least}. Motivated by this, we assign the LLM to simulate the role of users, allowing it to pose relevant questions $q_i$ pertaining to the entities within the initial query, which are used to subsequently generate fine-grained summaries highly relevant to the query. The LLM are required to ask the corresponding questions $q_i$ for each query entity $E_i$ around the query. The formalization is as follows:
    \begin{align}
        \text{Ask}(E_i, \mathcal{Q}) &= \{q_{i_1}, q_{i_2}, ... , q_{i_m}\}.
    \end{align}%
    \item \textbf{Summaries generation.} By retrieving in the graph, we gather a significant amount of fragmented information. When this large amount of coarse-grained information is fed directly into the LLM to generate the final response, it may be laced with noise that is not relevant to the query. Consequently, it becomes essential to consolidate this retrieved fragmented information into fine-grained information that is highly relevant to the query. For each query entity $E_i$, we instruct the LLM to consolidate the retrieved fragmented and relevant descriptions into a summary $\mathcal{S}_i$ that accurately reflects the information surrounding $E_i$, ensuring maximal relevance to the associated query. This step is based on the content retrieved for the query entity $E_i$ and takes into account any corresponding questions $q_i$ posed in the previous step. Finally, we instruct the LLM to use all the summaries to generate the final answer $\mathcal{S}$ for the query. The formalization is as follows:
    \begin{align}
        \mathcal{S}_i &= \text{Summarize}(\mathcal{G}'_i, \{q_{i_1}, q_{i_2}, ... , q_{i_m}\}), \\
        \mathcal{S} &= \text{Summarize}(\mathcal{Q}, \{\mathcal{S}_1, \mathcal{S}_2, ... , \mathcal{S}_n\}).
    \end{align}%
\end{itemize}

By applying the Query-Level Fine-Grained Summarization, FG-RAG effectively transforms the large amount of coarse-grained information in the graph into fine-grained information that is highly relevant to the query, reducing the impact of noise on the generation of the LLM.

\section{Experiments}

\begin{table*}[!t]
    \centering
    \renewcommand{\arraystretch}{0.8}
    \vspace{-0.1in}
    \resizebox{\linewidth}{!}{
    \begin{tabular}{lcccccccc}
        \toprule
        %\multirow{2}{*}{\textbf{QFS Tasks}} % 合并第一、二行的第一列，并填写内容
        & \multicolumn{2}{c}{\textbf{gpt-4o-mini}} 
        & \multicolumn{2}{c}{\textbf{qwen2.5-7b-instruct}} 
        & \multicolumn{2}{c}{\textbf{qwen2.5-3b-instruct}} 
        & \multicolumn{2}{c}{\textbf{qwen2.5-1.5b-instruct}} \\ 
        \cmidrule(lr){2-3} \cmidrule(lr){4-5} \cmidrule(lr){6-7} \cmidrule(lr){8-9}
        & NaiveRAG & \textbf{FG-RAG}
        & NaiveRAG & \textbf{FG-RAG}
        & NaiveRAG & \textbf{FG-RAG}
        & NaiveRAG & \textbf{FG-RAG} \\
        \midrule
        Comprehensiveness 
        & 35.52\%      & \textbf{64.48\%}     
        & 32.32\%      & \textbf{67.68\%}     
        & 24.96\%      & \textbf{75.04\%}     
        & 38.24\%      & \textbf{61.76\%}\\
        Diversity 
        & 29.20\%      & \textbf{70.80\%}     
        & 34.32\%      & \textbf{65.68\%}     
        & 21.44\%      & \textbf{78.56\%}     
        & 38.48\%      & \textbf{61.52\%}\\
        Empowerment 
        & 30.08\%      & \textbf{69.92\%}     
        & 33.12\%      & \textbf{66.88\%}     
        & 21.68\%      & \textbf{78.32\%}     
        & 37.68\%      & \textbf{62.32\%}\\
        Overall 
        & 30.96\%      & \textbf{69.04\%}     
        & 31.52\%      & \textbf{68.48\%}     
        & 21.76\%      & \textbf{78.24\%}     
        & 36.80\%      & \textbf{63.20\%}\\
        \cmidrule(lr){2-3} \cmidrule(lr){4-5} \cmidrule(lr){6-7} \cmidrule(lr){8-9}
        & LightRAG & \textbf{FG-RAG}
        & LightRAG & \textbf{FG-RAG}
        & LightRAG & \textbf{FG-RAG}
        & LightRAG & \textbf{FG-RAG} \\
        \midrule
        Comprehensiveness 
        & 43.68\%      & \textbf{56.32\%}     
        & 37.76\%      & \textbf{62.24\%}     
        & 39.44\%      & \textbf{60.56\%}     
        & 35.28\%      & \textbf{64.72\%}\\
        Diversity 
        & 42.72\%      & \textbf{57.28\%}     
        & 46.32\%      & \textbf{53.68\%}     
        & 38.48\%      & \textbf{61.52\%}     
        & 36.48\%      & \textbf{63.52\%}\\
        Empowerment 
        & 40.96\%      & \textbf{59.04\%}     
        & 35.12\%      & \textbf{64.88\%}     
        & 34.48\%      & \textbf{65.52\%}     
        & 31.68\%      & \textbf{68.32\%}\\
        Overall 
        & 41.36\%      & \textbf{58.64\%}     
        & 36.00\%      & \textbf{64.00\%}     
        & 36.96\%      & \textbf{63.04\%}     
        & 33.76\%      & \textbf{66.24\%}\\
        \cmidrule(lr){2-3} \cmidrule(lr){4-5} \cmidrule(lr){6-7} \cmidrule(lr){8-9}
        & GraphRAG & \textbf{FG-RAG}
        & GraphRAG & \textbf{FG-RAG}
        & GraphRAG & \textbf{FG-RAG}
        & GraphRAG & \textbf{FG-RAG} \\
        \midrule
        Comprehensiveness 
        & 39.60\%      & \textbf{60.40\%}     
        & 24.32\%      & \textbf{75.68\%}     
        & 23.24\%      & \textbf{76.76\%}     
        & 22.00\%      & \textbf{78.00\%}\\
        Diversity 
        & 34.48\%      & \textbf{65.52\%}     
        & 32.96\%      & \textbf{67.04\%}     
        & 31.12\%      & \textbf{68.88\%}     
        & 35.12\%      & \textbf{64.88\%}\\
        Empowerment 
        & 33.76\%      & \textbf{66.24\%}     
        & 18.56\%      & \textbf{81.44\%}     
        & 16.40\%      & \textbf{83.60\%}     
        & 19.36\%      & \textbf{80.64\%}\\
        Overall 
        & 35.36\%      & \textbf{64.64\%}     
        & 22.00\%      & \textbf{78.00\%}     
        & 19.28\%      & \textbf{80.72\%}     
        & 22.00\%      & \textbf{78.00\%}\\
        \bottomrule
    \end{tabular}
    }
    \vspace{-0.1in}
    \caption{Average win rates of four evaluation metrics for four LLMs when solving the QFS task across five datasets. The comparison is made between baselines and FG-RAG.}
    \label{tab:QFS}
    \vspace{-0.1in}
\end{table*}

\subsection{Experimental Settings}

To comprehensively analyze FG-RAG, we compare it against other RAG methods on both the QFS task and multi-hop QA task.

\noindent\textbf{Datasets}\quad We select four specialized datasets from the UltraDomain benchmark~\cite{memorag}, covering the domains of Agriculture, Art, Legal, and a Mixed category. Furthermore, we include a benchmark dataset comprising news articles published between September 2013 and December 2023~\cite{multihop}. The size of each dataset ranges from 600,000 to 5,000,000 tokens. To generate the queries for the QFS task, we compile all textual data from each dataset to serve as context, following the method described by GraphRAG~\cite{graphrag}. Specifically, we instruct the LLM to create 5 RAG users, each with 5 unique tasks. For every user-task pair, the LLM generates 5 queries that require a thorough understanding of the entire corpus. As a result, a total of 125 queries are generated per dataset.

While the primary focus is on solving the QFS task, FG-RAG's adaptability is also tested by concurrently conducting experiments in the multi-hop QA task. Using the same dataset of news articles as for the QFS task, we randomly select 125 queries from each of the three multi-hop query categories: reference, comparison, and temporal queries~\cite{multihop}. To answer these queries, reasoning must be applied across 2 to 4 documents within the dataset.

\begin{table}
    \centering
    \renewcommand{\arraystretch}{0.8}
    \vspace{-0.1in}
    \resizebox{\linewidth}{!}{
    \begin{tabular}{ccccc}
        \toprule
        & gpt-4o-mini & Q-7B-it & Q-3B-it & Q-1.5B-it \\
        \midrule
        NaiveRAG    & 54.40\%          & 43.20\%        & \textbf{31.73\%}        & 36.80\%\\
        LightRAG     & 43.47\%          & 37.60\%        & 29.60\%        & 37.60\%\\
        GraphRAG     & 65.33\%          & 34.93\%        & 24.00\%        & \textbf{50.13\%}\\
        FG-RAG     & \textbf{69.87\%}          & \textbf{57.33\%}        & 30.13\%        & 39.20\%\\
        \bottomrule
    \end{tabular}
    }
    \vspace{-0.1in}
    \caption{Baselines and FG-RAG accuracy rates in multi-hop QA.}
    \label{tab:multihop}
    \vspace{-0.1in}
\end{table}

\noindent\textbf{Baselines}\quad FG-RAG is compared against the following state-of-the-art methods across all datasets.
\begin{itemize}
    \item NaiveRAG~\cite{naiverag}: This method segments text into chunks, stores them as embeddings in a vector database, and retrieves the most similar chunks during queries for efficient matching.
    \item GraphRAG~\cite{graphrag}: This graph-enhanced RAG system generates descriptions for elements and produces community reports after aggregating the nodes into communities, responding to global queries by retrieving the most relevant community information.
    \item LightRAG~\cite{lightrag}: This graph-enhanced RAG system combines graph structure with vector representation and utilizes a dual-level retrieval paradigm to improve retrieval efficiency and retrieval relevance.
\end{itemize}

\begin{table*}[!t]
    \centering
    \renewcommand{\arraystretch}{0.8}
    \vspace{-0.1in}
    \resizebox{\linewidth}{!}{
    \begin{tabular}{lcccccccc}
        \toprule
        %\multirow{2}{*}{\textbf{QFS Tasks}} % 合并第一、二行的第一列，并填写内容
        & \multicolumn{2}{c}{\textbf{gpt-4o-mini}} 
        & \multicolumn{2}{c}{\textbf{qwen2.5-7b-instruct}} 
        & \multicolumn{2}{c}{\textbf{qwen2.5-3b-instruct}} 
        & \multicolumn{2}{c}{\textbf{qwen2.5-1.5b-instruct}} \\ 
        \cmidrule(lr){2-3} \cmidrule(lr){4-5} \cmidrule(lr){6-7} \cmidrule(lr){8-9}
        & NaiveRAG & \textbf{FG-RAG}
        & NaiveRAG & \textbf{FG-RAG}
        & NaiveRAG & \textbf{FG-RAG}
        & NaiveRAG & \textbf{FG-RAG} \\
        \midrule
        Comprehensiveness 
        & 35.52\%      & \textbf{64.48\%}     
        & 32.32\%      & \textbf{67.68\%}     
        & 24.96\%      & \textbf{75.04\%}     
        & 38.24\%      & \textbf{61.76\%}\\
        Diversity 
        & 29.20\%      & \textbf{70.80\%}     
        & 34.32\%      & \textbf{65.68\%}     
        & 21.44\%      & \textbf{78.56\%}     
        & 38.48\%      & \textbf{61.52\%}\\
        Empowerment 
        & 30.08\%      & \textbf{69.92\%}     
        & 33.12\%      & \textbf{66.88\%}     
        & 21.68\%      & \textbf{78.32\%}     
        & 37.68\%      & \textbf{62.32\%}\\
        Overall 
        & 30.96\%      & \textbf{69.04\%}     
        & 31.52\%      & \textbf{68.48\%}     
        & 21.76\%      & \textbf{78.24\%}     
        & 36.80\%      & \textbf{63.20\%}\\
        \cmidrule(lr){2-3} \cmidrule(lr){4-5} \cmidrule(lr){6-7} \cmidrule(lr){8-9}
        & NaiveRAG & \textbf{w/o CAEE}
        & NaiveRAG & \textbf{w/o CAEE}
        & NaiveRAG & \textbf{w/o CAEE}
        & NaiveRAG & \textbf{w/o CAEE} \\
        \midrule
        Comprehensiveness 
        & 37.76\%      & \textbf{62.24\%}     
        & 37.20\%      & \textbf{62.80\%}     
        & 29.84\%      & \textbf{70.16\%}     
        & 41.04\%      & \textbf{58.96\%}\\
        Diversity 
        & 34.08\%      & \textbf{65.92\%}     
        & 38.00\%      & \textbf{62.00\%}     
        & 28.24\%      & \textbf{71.76\%}     
        & 42.80\%      & \textbf{57.20\%}\\
        Empowerment 
        & 32.32\%      & \textbf{67.68\%}     
        & 35.20\%      & \textbf{64.80\%}     
        & 25.92\%      & \textbf{74.08\%}     
        & 39.68\%      & \textbf{60.32\%}\\
        Overall 
        & 32.24\%      & \textbf{67.76\%}     
        & 35.76\%      & \textbf{64.24\%}     
        & 25.76\%      & \textbf{74.24\%}     
        & 39.44\%      & \textbf{60.56\%}\\
        \cmidrule(lr){2-3} \cmidrule(lr){4-5} \cmidrule(lr){6-7} \cmidrule(lr){8-9}
        & NaiveRAG & \textbf{w/o QLFGS}
        & NaiveRAG & \textbf{w/o QLFGS}
        & NaiveRAG & \textbf{w/o QLFGS}
        & NaiveRAG & \textbf{w/o QLFGS} \\
        \midrule
        Comprehensiveness 
        & 48.48\%      & \textbf{51.52\%}     
        & 44.64\%      & \textbf{55.36\%}     
        & 41.92\%      & \textbf{58.08\%}     
        & 47.68\%      & \textbf{52.32\%}\\
        Diversity 
        & 42.00\%      & \textbf{58.00\%}     
        & 39.76\%      & \textbf{60.24\%}     
        & 33.36\%      & \textbf{66.64\%}     
        & 43.20\%      & \textbf{56.80\%}\\
        Empowerment 
        & 46.88\%      & \textbf{53.12\%}
        & 45.04\%      & \textbf{54.96\%}     
        & 42.40\%      & \textbf{57.60\%}     
        & 49.76\%      & \textbf{50.24\%}\\
        Overall 
        & 47.12\%      & \textbf{52.88\%}     
        & 43.60\%      & \textbf{56.40\%}     
        & 40.80\%      & \textbf{59.20\%}     
        & 47.92\%      & \textbf{52.08\%}\\
        \bottomrule
    \end{tabular}
    }
    \vspace{-0.1in}
    \caption{Average win rates of FG-RAG compared to the two ablation models on four evaluation metrics. CAEE refers to the graph retrieval with Context-Aware Entity Expansion and QLFGS refers to Query-Level Fine-Grained Summarization.}
    \label{tab:ablation}
    \vspace{-0.1in}
\end{table*}

\noindent\textbf{Evaluation Metrics}\quad The following are the evaluation metrics for the QFS task and the multi-hop QA task:
\begin{itemize}
    \item \textbf{QFS task}: Following LightRAG~\cite{lightrag}, we consider the following metrics to evaluate the answer to queries in the QFS task: 1) \textbf{Comprehensiveness}: This metric is used to assess the thoroughness of the answer in addressing all aspects and details of the query. 2) \textbf{Diversity}: This metric is used to assess the variety and richness of the answer in terms of providing different perspectives and insights pertinent to the query. 3) \textbf{Empowerment}: This metric is used to assess the effectiveness of the answer in terms of helping the reader to understand the topic and make informed judgments. 4) \textbf{Overall}: This metric is used to assess the overall performance of the answer based on the cumulative performance of the three preceding metrics.
    \item \textbf{Multi-hop QA task}: In the multi-hop QA task, we require each baseline to output a word or entity as an answer if the information in the retrieved content is sufficient. When assessing the model's responses, we compare them against the standard answers, and they must match exactly to be deemed correct.
\end{itemize}

\subsection{Overall Performance on the QFS Task}

In this study, we evaluate FG-RAG in comparison with each baseline using four distinct evaluation metrics, across four LLMs, and on five diverse datasets. Due to space constraints, we provide detailed data on each dataset and compare it with FastGraphRAG in the supplementary material. The findings are summarized in Table ~\ref{tab:QFS}, from which we derive the subsequent conclusions:

\begin{itemize}
    \item \textbf{FG-RAG outperforms other baselines in all metrics.} FG-RAG demonstrates an enhancement in all evaluation metrics when contrasted with alternative baselines. Among GraphRAG-based methods applied to the QFS task, LightRAG particularly stands out due to its high diversity. However, FG-RAG not only improves upon all other metrics but also further enhances diversity when compared to LightRAG. We attribute this advantage to the ability of FG-RAG to retrieve a wide range of relevant contexts for the query, facilitating the generation of summaries and refining the granularity of the information retrieved.
    \item \textbf{FG-RAG reduces the reliance on the comprehension abilities of the LLM.} When the LLM used is changed to a model with fewer parameters, such as qwen2.5-3b-instruct, the win rates of FG-RAG for all evaluation metrics is further improved significantly compared to the use of gpt-4o-mini. When using qwen2.5-1.5b-instruct, other GraphRAG-based methods have degraded to be comparable to or even worse than NaiveRAG on various metrics, and our FG-RAG still maintains a clear advantage. This improvement can be attributed to the design of Query-Level Fine-Grained Summarization, which focuses on consolidating the large amount of coarse-grained information in the graph into fine-grained information that is highly relevant to the query, thus relying less on the comprehension capabilities of the LLM.
\end{itemize}

\begin{table*}[!t]
\centering
\vspace{-0.1in}
\begin{tabular}{|c|p{14cm}|}
\hline
\textbf{Query} & How have artistic crossovers influenced political and social relations historically? \\ \hline
\textbf{FG-RAG} & ... The collaboration between diverse creative expressions has historically fostered dialogue around contemporary political climates and cultural dynamics. 
1. \textbf{Artistic Expression as a Catalyst for Change} [...]  
2. \textbf{Public Art and Community Engagement} [...]  
3. \textbf{The Role of Historical Figures} [...]  
4. \textbf{Gender and Cultural Representation} [...] 
\textbf{Conclusion}: In conclusion, the historical interplay between artistic crossovers and political/social relations ... \\ \hline
\textbf{LightRAG} &
... Below are some of the most informative metrics commonly used in this context, derived from the data and concepts surrounding recommendation systems.  
1. \textbf{Cultural Exchange and Its Political Implications} [...]  
2. \textbf{Art as a Tool for Social Movements} [...]  
3. \textbf{Reflecting and Shaping Societal Values} [...]  
\textbf{Conclusion}: Artistic crossovers have historically played a significant role in ... \\ \hline
\textbf{LLM Decision} & \textbf{Comprehensiveness}:  
``Winner": ``Answer 1 (FG-RAG)", % \newline

``Explanation": ``Answer 1 provides a thorough examination of the role of artistic crossovers in political and social relations, encompassing various aspects such as historical examples, the ..."

\textbf{Diversity}:  
``Winner": ``Answer 1 (FG-RAG)", %

``Explanation": ``Answer 1 offers a greater variety of perspectives on the topic. It discusses not only the impact of individual artists but also artistic collaborations, community engagement ..."

\textbf{Empowerment}:  
``Winner": ``Answer 1 (FG-RAG)", % \newline

``Explanation": ``Answer 1 empowers readers by providing them with a nuanced understanding of the subject, illustrating how art serves both as a mirror and a catalyst for societal change ..."

\textbf{Overall Winner}:  
``Winner": ``Answer 1 (FG-RAG)",  %\newline

``Explanation": ``Answer 1 is the overall winner as it excels in all three criteria: it is comprehensive in addressing various dimensions of the question, diverse in the perspectives and 
..." \\ \hline

\end{tabular}
\vspace{-0.1in}
\caption{A representative case comparing LightRAG to our FG-RAG method.}
\label{tab:case}
\vspace{-0.1in}
\end{table*}

\subsection{Overall Performance on the Multi-hop QA Task}

LightRAG, GraphRAG, and FG-RAG are primarily designed for the QFS task with extensive text data, emphasizing data integration over detailed reasoning about the retrieved information. Queries necessitating inference across a small set of documents to derive an answer pose a significant challenge for all aforementioned methods. In order to show the better generalization ability of FG-RAG compared to other Graph-based methods used to solve the QFS task, we evaluate the accuracy of FG-RAG compared to each baseline in the multi-hop QA task(Table ~\ref{tab:multihop}), leading to the following conclusions:

Compared to other RAG systems, our FG-RAG model has demonstrated substantial improvements in accuracy across most models, with peak performance on the gpt-4o-mini and qwen2.5-7b-instruct models. This can be attributed to the fact that FG-RAG is query-aware and reduces the granularity of the retrieved content.

\subsection{Ablation Studies}

To further investigate the impact of Query-Level Fine-Grained Summarization and the graph retrieval with Context-Enhanced Entity Expansion, we conduct ablation studies. We compare two ablated models against FG-RAG across all five datasets. Based on the experimental results in Table ~\ref{tab:ablation}, the following conclusions can be drawn:

\begin{itemize}
    \item \textbf{Effectiveness of the graph retrieval with Context-Aware Entity Expansion.} The results indicate that not using the graph retrieval with Context-Aware Entity Expansion yields a reduction in all metrics across multiple models and datasets when contrasted with FG-RAG, particularly in terms of answer diversity. 
    These findings illustrate that the graph retrieval with Context-Aware Entity Expansion enables more contextual information to be retrieved from the graph, while ensuring that the retrieved information is reasonably relevant to the query.
    \item \textbf{Effectiveness of Query-Level Fine-Grained Summarization.} The results indicate that, compared to FG-RAG, the performance across all four metrics significantly decreases when Query-Level Fine-Grained Summarization is not used. These findings demonstrate that Query-Level Fine-Grained Summarization constitutes the core component of FG-RAG, which effectively provides enough fine-grained information highly relevant to the query to solve the QFS task.
\end{itemize}

\subsection{Case Study}

In order to more intuitively compare the advantages of FG-RAG over other RAG systems in solving the QFS task, we use FG-RAG and other baselines separately to address a complex query about how artistic crossovers affect political and social relations. Due to space constraints, we only present a comparison between FG-RAG and LightRAG. Additional case studies can be available in the supplementary material.
Table~\ref{tab:case} shows that FG-RAG outperforms LightRAG in all four metrics, with detailed assessments for each metric listed below:

\begin{itemize}
    \item \textbf{Comprehensiveness}: FG-RAG offers an extensive examination of how art influences political and social relations across borders, exploring a broad spectrum of related themes. 
    FG-RAG achieves this by using Context-Aware Entity Expansion in the process of graph retrieval, which enriches contextual information to expand more entities during retrieval, making the results more comprehensive.
    \item \textbf{Diversity}: FG-RAG's response is significantly richer, highlighting the diverse ways in which the arts influence global society.
    This effectiveness comes from the fact that FG-RAG decomposes the query into multiple entities and applies Context-Aware Entity Expansion during the retrieval process for each entity within the graph. Consequently, it furnishes the LLM with retrieval information from various perspectives and insights.
    \item \textbf{Empowerment}: FG-RAG not only empowers users by providing them with a nuanced understanding of the topic but also enriches their comprehension by offering numerous pertinent examples. 
    FG-RAG enables the LLM to formulate questions around the query entities from the perspective of the user and filter out extraneous noise through fine-grained summarization, thereby making the answers generated by the LLM more understandable.
    \item \textbf{Overall}: FG-RAG offers an in-depth analysis of the query, blending various viewpoints and examples to illuminate the complex relationship between artistic crossover and political or social engagement.
\end{itemize}

\subsection{Comparison of Method Efficiency}

\begin{table}[!t]
\centering
\vspace{-0.1in}
\resizebox{\linewidth}{!}{
\begin{tabular}{|c|c|c|c|c|}
\hline

Metric & NaiveRAG & GraphRAG & LightRAG & FG-RAG \\ \hline
Graph Indexing Tokens & 0 & 7,707,566 & 5,111,680 & \textbf{4,863,801} \\ \hline
Query Tokens & 1,377,432 & 36,593,256 & 2,884,462 & \textbf{1,356,951} \\ \hline
\end{tabular}
}
\vspace{-0.1in}
\caption{Tokens of FG-RAG and baselines on the Mix dataset.}
\label{tab:cost}
\vspace{-0.1in}
\end{table}

We compare the number of tokens consumed by FG-RAG and baselines when indexing the graph and responding to queries for the QFS task on the Mix dataset. The dataset contains approximately 600,000 tokens and a total of 125 queries. The results are presented in Table ~\ref{tab:cost}.

Compared to GraphRAG, both LightRAG and FG-RAG significantly reduce the number of tokens that are used when indexing the graph and responding to queries. Since the graph indexing process of FG-RAG is similar to that of LightRAG, FG-RAG consumes only slightly fewer tokens\footnote{FG-RAG uses fewer examples in the prompt
compared to LightRAG for extracting entities and relationships.} than LightRAG when indexing a new dataset into the graph. FG-RAG processes the entities in the query at a fine-grained level, which can significantly decrease token overhead while improving generation quality. In response to queries, FG-RAG is more economical than LightRAG, reducing token usage by more than 50\%. Although NaiveRAG uses a comparable number of tokens as FG-RAG in response to queries, its generation quality is significantly inferior to FG-RAG when solving the QFS task.

\section{Conclusion}

In this work, we propose FG-RAG, an innovative RAG framework designed to handle the QFS task through the integration of the graph retrieval with Context-Aware Entity Expansion and Query-Level Fine-Grained Summarization.
Our framework effectively converts the large amount of coarse-grained information in the knowledge graph from external documents into fine-grained information that is highly relevant to the query, and incorporates more contextual information related to the query entity during the retrieval process, thus improving the quality of the response in solving the QFS task.

\bibliographystyle{named}
\bibliography{main}

\clearpage

\section{Supplementary Materials}
\subsection{Detailed Experiment Settings}

\begin{table}[H]
\centering
\resizebox{\linewidth}{!}{
    \begin{tabular}{lcccccc}
    \toprule
    \textbf{Statistics}  & \textbf{Agriculture} & \textbf{Art} & \textbf{Legal} & \textbf{Mix} & \textbf{News}\\
    \midrule
    Total Tokens    & 1,923,158 & 3,581,243 & 4,719,597 & 602,561 & 1,409,743\\
    Total Chunks    & 1,838 & 3,414 & 5,424 & 596 & 1,614 \\
    \bottomrule
    \end{tabular}
}
\caption{The detailed statistics of the five datasets.}
\label{tab:chunk}
\end{table}

\noindent\textbf{Datasets Statistics}\quad In the QFS task, we choose a total of five datasets for evaluating the FG-RAG. The total number of tokens and the number of chunks partitioned into each dataset are summarized in Table \ref{tab:chunk}.

In the multi-hop QA task, we choose the News dataset with multi-hop QA. Given the extensive collection of QA pairs in the multi-hop QA, we opted for a random selection of 125 pairs from each category——namely, inference, comparison, and temporal queries—-for our evaluation process. The details of the three types of queries are described below:

\begin{itemize}
    \item \textbf{Inference query} These queries are developed by integrating the various descriptions of the bridge-entity from multiple claims, ultimately leading to the identification of the entity itself.
    \item \textbf{Comparison query} These queries are crafted to highlight the similarities and differences concerning the bridge entity or topic, with the answer usually being a clear ``yes” or ``no” based on the comparative analysis.
    \item \textbf{Temporal query} These queries examine the sequence of events at different points in time, with answers typically being a ``yes" or ``no", or a single temporal indicator such as ``before" or ``after".
\end{itemize}

\noindent\textbf{Implementation Details}\quad The chunk size for all datasets is standardized at 1200 tokens, with a 100-token overlap between consecutive chunks. In addition, the gleaning parameter is fixed to 1 for all GraphRAG-based methods that need to extract entities and relationships from the dataset.

\noindent\textbf{Evaluation Details}\quad In the QFS task, we use the evaluation method presented in LightRAG~\cite{lightrag} to compare the answers given by different methods, and the LLM used in the evaluation process is gpt-4o-mini. We alternate the placement of the two answers in the prompt when comparing the answers given by two methods and calculate win rates accordingly. Specifically, we will treat the answer given by a method as Answer 1 and Answer 2 respectively at one time, thus ensuring fair comparisons and minimizing possible bias due to the order in the prompt. In the multi-hop QA task, we judge an answer to be correct when and only when it is identical to the standard answer ignoring case.

\subsection{Detailed Experiment Results}

In this paper, when assessing the performance of the QFS task, we initially display the average performance of each model across all datasets. To gain a more detailed understanding of how the performance of FG-RAG compares to other baselines across various datasets, in Table~\ref{tab:detail} we present specific experimental results for each dataset. In addition to the baselines mentioned in the paper, we also made a comparison with \href{https://github.com/circlemind-ai/fast-graphrag}{FastGraphRAG}. It is a graph-enhanced RAG system integrates graph structure with vector representation, utilizing the PageRank algorithm for graph retrieval to enhance the accuracy and reliability of information retrieval. However, due to the outputs of certain LLMs, like qwen2.5-7b-instruct, failing to meet FastGraphRAG's strict formatting requirements, we are only able to get its response to the query when using gpt-4o-mini.

\subsection{Overview of the Prompts Used in FG-RAG}

\subsubsection{Prompts for query entities extraction}

\begin{figure*}[!t]
    \centering
    \includegraphics[width=\textwidth]{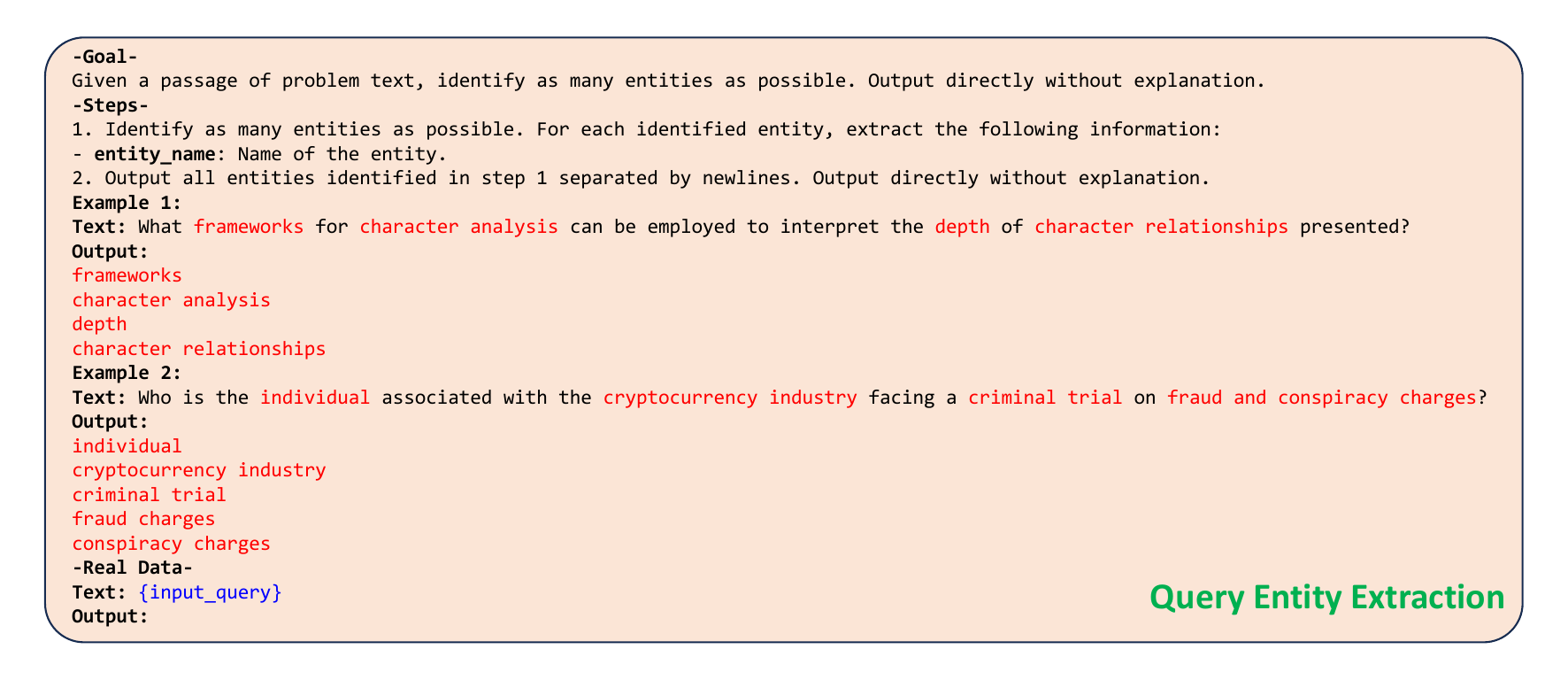}
    \caption{Prompts for query entities extraction.}
    \label{fig:prompt1}
\end{figure*}

The prompt in Figure~\ref{fig:prompt1} instructs the LLM to extract entities from the query. To prevent the LLM from overlooking key entities in queries during entity extraction, we apply few-shot prompting.

\subsubsection{Prompts for relevant questions formulation}

\begin{figure*}
    \centering
    \includegraphics[width=1\textwidth]{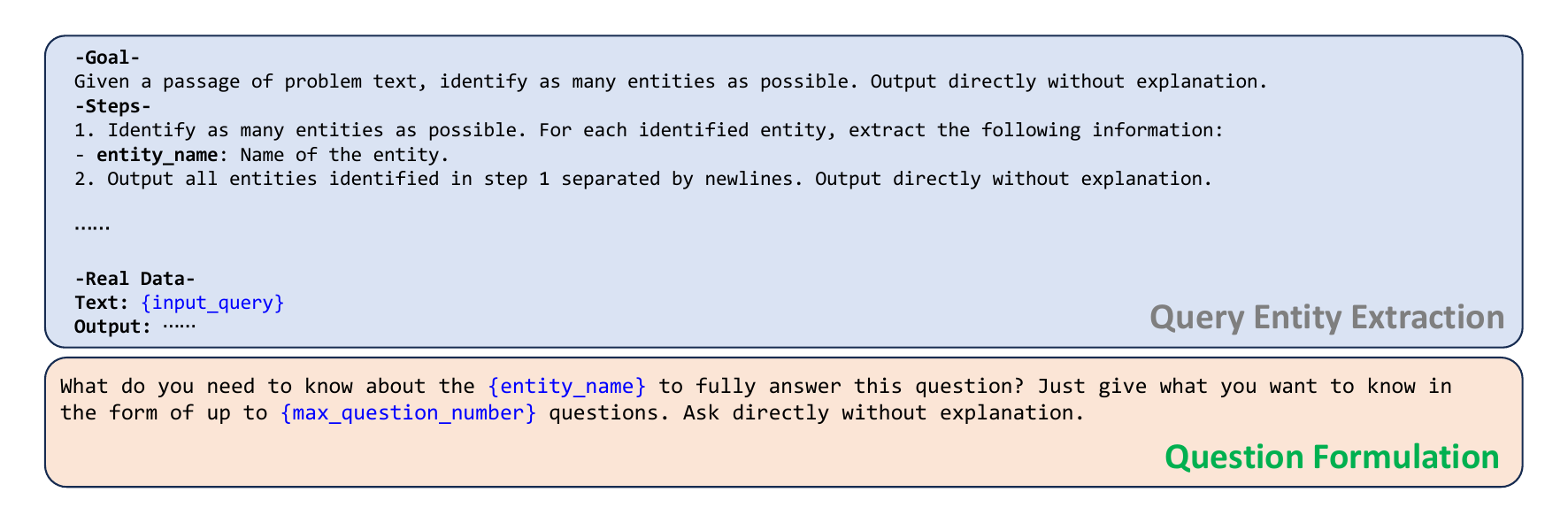}
    \caption{Prompts for relevant questions formulation.}
    \label{fig:prompt2}
\end{figure*}

The prompt in Figure~\ref{fig:prompt2} instructs the LLM to ask relevant questions around a particular entity. Assuming that the query entity extracted in the entities extraction phase are $\{E_1, E_2, ... , E_n\}$, we direct the LLM to ask a certain number of questions around the query entity $E_i$ based on the context of the query entity extraction step.

\subsubsection{Prompts for summaries generation}

\begin{figure*}
    \centering
    \includegraphics[width=1\textwidth]{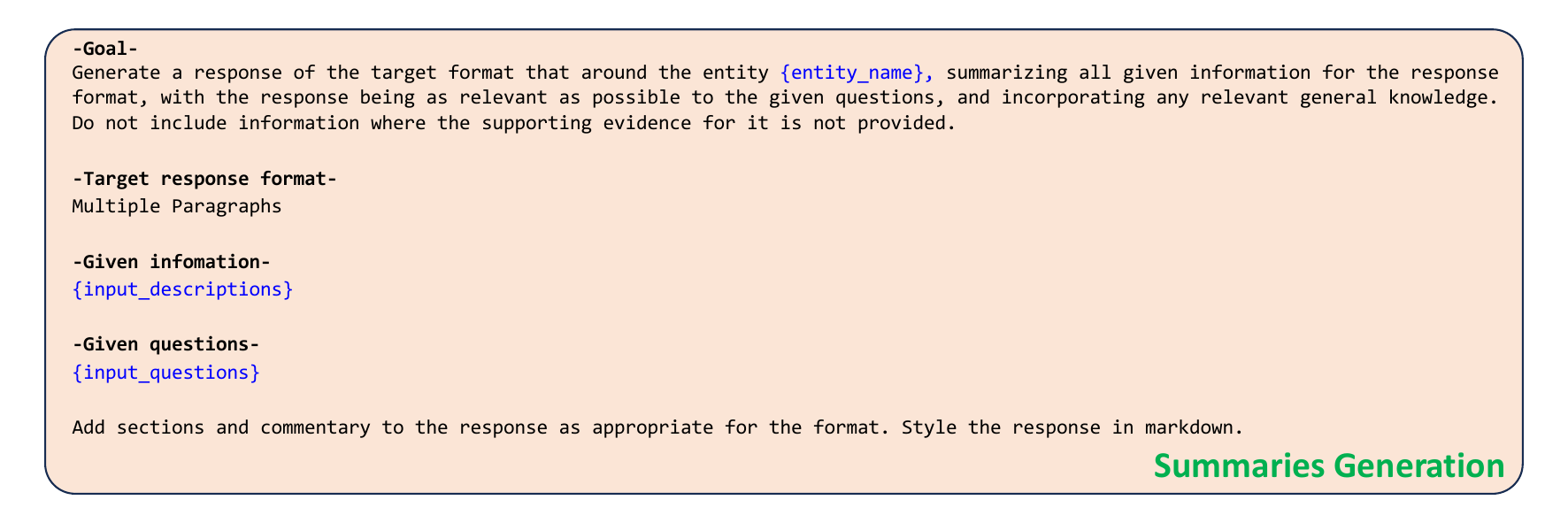}
    \caption{Prompts for summaries generation.}
    \label{fig:prompt3}
\end{figure*}

The prompt in Figure~\ref{fig:prompt3} instructs the LLM to generate a summary based on the retrieved content around a particular entity and some questions related to the entity. The generated summary consists of multiple paragraphs and is output as markdown in sections.

\subsubsection{Other prompts}

Prompts used in other steps such as extracting entities and their relationships from text chunks, QFS task question generation, and evaluation of answer pairs are similar to those used in LightRAG and are not repeated here.

\subsection{Case Studies Comparing FG-RAG with Other Baselines}

In order to visualize the superiority of FG-RAG over other baselines in terms of comprehensiveness, diversity, and empowerment, we present some case studies in addition to LightRAG. Table~\ref{tab:naive}, Table~\ref{tab:graph}, and Table~\ref{tab:fastgraph} show the answers given by FG-RAG compared to NaiveRAG, GraphRAG, and \href{https://github.com/circlemind-ai/fast-graphrag}{FastGraphRAG} in solving a complex query about how artistic crossovers affect political and social relations, respectively.

\clearpage
\newpage
\begin{table*}
    \centering
    \renewcommand{\arraystretch}{0.8}
    \resizebox{\linewidth}{!}{
    \begin{tabular}{lcccccccccc}
        \toprule
        \multirow{2}{*}{\textbf{gpt-4o-mini}}
        & \multicolumn{2}{c}{\textbf{Agriculture}} 
        & \multicolumn{2}{c}{\textbf{Art}} 
        & \multicolumn{2}{c}{\textbf{CS}} 
        & \multicolumn{2}{c}{\textbf{Mix}}
        & \multicolumn{2}{c}{\textbf{News}} \\ 
        \cmidrule(lr){2-3} \cmidrule(lr){4-5} \cmidrule(lr){6-7} \cmidrule(lr){8-9} \cmidrule(lr){10-11} 
        & NaiveRAG & \textbf{FG-RAG}
        & NaiveRAG & \textbf{FG-RAG}
        & NaiveRAG & \textbf{FG-RAG}
        & NaiveRAG & \textbf{FG-RAG} 
        & NaiveRAG & \textbf{FG-RAG} \\
        \midrule
        Comprehensiveness 
        & 37.2\%      & \textbf{62.8\%}     
        & 36.0\%      & \textbf{64.0\%}     
        & 33.2\%      & \textbf{66.8\%}
        & 33.2\%      & \textbf{66.8\%}
        & 38.0\%      & \textbf{62.0\%}\\
        Diversity 
        & 32.0\%      & \textbf{68.0\%}     
        & 33.2\%      & \textbf{66.8\%}     
        & 11.2\%      & \textbf{88.8\%}
        & 37.6\%      & \textbf{62.4\%}
        & 32.0\%      & \textbf{68.0\%}\\
        Empowerment 
        & 34.4\%      & \textbf{65.6\%}     
        & 32.4\%      & \textbf{67.6\%}     
        & 21.6\%      & \textbf{78.4\%}
        & 30.4\%      & \textbf{69.6\%}
        & 31.6\%      & \textbf{68.4\%}\\
        Overall 
        & 36.0\%      & \textbf{64.0\%}     
        & 34.4\%      & \textbf{65.6\%}     
        & 23.2\%      & \textbf{76.8\%}
        & 28.8\%      & \textbf{71.2\%}
        & 32.4\%      & \textbf{67.6\%}\\
        \cmidrule(lr){2-3} \cmidrule(lr){4-5} \cmidrule(lr){6-7} \cmidrule(lr){8-9} \cmidrule(lr){10-11} 
        & LightRAG & \textbf{FG-RAG}
        & LightRAG & \textbf{FG-RAG}
        & LightRAG & \textbf{FG-RAG}
        & LightRAG & \textbf{FG-RAG}
        & LightRAG & \textbf{FG-RAG} \\
        \midrule
        Comprehensiveness 
        & 46.0\%      & \textbf{54.0\%}     
        & 40.4\%      & \textbf{59.6\%}     
        & 47.6\%      & \textbf{52.4\%}
        & 39.2\%      & \textbf{60.8\%}
        & 45.2\%      & \textbf{54.8\%}\\
        Diversity 
        & 46.4\%      & \textbf{53.6\%}     
        & 40.0\%      & \textbf{60.0\%}     
        & 38.4\%      & \textbf{61.6\%}
        & 45.6\%      & \textbf{54.4\%}
        & 43.2\%      & \textbf{56.8\%}\\
        Empowerment 
        & 42.0\%      & \textbf{58.0\%}     
        & 40.8\%      & \textbf{59.2\%}     
        & 45.2\%      & \textbf{54.8\%}
        & 35.6\%      & \textbf{64.4\%}
        & 41.2\%      & \textbf{58.8\%}\\
        Overall 
        & 42.0\%      & \textbf{58.0\%}     
        & 40.0\%      & \textbf{60.0\%}     
        & 44.8\%      & \textbf{55.2\%}
        & 36.0\%      & \textbf{64.0\%}
        & 44.0\%      & \textbf{56.0\%}\\
        \cmidrule(lr){2-3} \cmidrule(lr){4-5} \cmidrule(lr){6-7} \cmidrule(lr){8-9} \cmidrule(lr){10-11} 
        & GraphRAG & \textbf{FG-RAG}
        & GraphRAG & \textbf{FG-RAG}
        & GraphRAG & \textbf{FG-RAG}
        & GraphRAG & \textbf{FG-RAG}
        & GraphRAG & \textbf{FG-RAG} \\
        \midrule
        Comprehensiveness 
        & 39.2\%      & \textbf{60.8\%}     
        & 36.4\%      & \textbf{63.6\%}     
        & 40.8\%      & \textbf{59.2\%}
        & 41.2\%      & \textbf{58.8\%}
        & 40.4\%      & \textbf{59.6\%}\\
        Diversity 
        & 30.8\%      & \textbf{69.2\%}     
        & 30.8\%      & \textbf{69.2\%}     
        & 32.4\%      & \textbf{67.6\%}
        & 42.4\%      & \textbf{57.6\%}
        & 36.0\%      & \textbf{64.0\%}\\
        Empowerment 
        & 34.0\%      & \textbf{66.0\%}     
        & 30.4\%      & \textbf{69.6\%}     
        & 30.0\%      & \textbf{70.0\%}
        & 33.6\%      & \textbf{66.4\%}
        & 40.8\%      & \textbf{59.2\%}\\
        Overall 
        & 34.8\%      & \textbf{65.2\%}     
        & 32.8\%      & \textbf{67.2\%}     
        & 34.4\%      & \textbf{65.6\%}
        & 35.2\%      & \textbf{64.8\%}
        & 39.6\%      & \textbf{60.4\%}\\
        \cmidrule(lr){2-3} \cmidrule(lr){4-5} \cmidrule(lr){6-7} \cmidrule(lr){8-9} \cmidrule(lr){10-11} 
        & FastGraphRAG & \textbf{FG-RAG}
        & FastGraphRAG & \textbf{FG-RAG}
        & FastGraphRAG & \textbf{FG-RAG}
        & FastGraphRAG & \textbf{FG-RAG}
        & FastGraphRAG & \textbf{FG-RAG} \\
        \midrule
        Comprehensiveness 
        & 36.4\%      & \textbf{63.6\%}     
        & 31.2\%      & \textbf{68.8\%}     
        & 26.4\%      & \textbf{73.6\%}
        & 32.0\%      & \textbf{68.0\%}
        & 35.2\%      & \textbf{64.8\%}\\
        Diversity 
        & 30.4\%      & \textbf{69.6\%}     
        & 27.6\%      & \textbf{72.4\%}     
        & 24.4\%      & \textbf{75.6\%}
        & 30.8\%      & \textbf{69.2\%}
        & 33.6\%      & \textbf{66.4\%}\\
        Empowerment 
        & 32.8\%      & \textbf{67.2\%}     
        & 27.6\%      & \textbf{72.4\%}     
        & 21.2\%      & \textbf{78.8\%}
        & 27.6\%      & \textbf{72.4\%}
        & 28.4\%      & \textbf{71.6\%}\\
        Overall 
        & 33.2\%      & \textbf{66.8\%}     
        & 28.8\%      & \textbf{71.2\%}     
        & 22.4\%      & \textbf{77.6\%}
        & 27.2\%      & \textbf{72.8\%}
        & 30.4\%      & \textbf{69.6\%}\\
        \midrule
        \multirow{2}{*}{\textbf{qwen2.5-7b-instruct}} % 合并第一、二行的第一列，并填写内容
        & \multicolumn{2}{c}{\textbf{Agriculture}} 
        & \multicolumn{2}{c}{\textbf{Art}} 
        & \multicolumn{2}{c}{\textbf{CS}} 
        & \multicolumn{2}{c}{\textbf{Mix}}
        & \multicolumn{2}{c}{\textbf{News}} \\ 
        \cmidrule(lr){2-3} \cmidrule(lr){4-5} \cmidrule(lr){6-7} \cmidrule(lr){8-9} \cmidrule(lr){10-11} 
        & NaiveRAG & \textbf{FG-RAG}
        & NaiveRAG & \textbf{FG-RAG}
        & NaiveRAG & \textbf{FG-RAG}
        & NaiveRAG & \textbf{FG-RAG} 
        & NaiveRAG & \textbf{FG-RAG} \\
        \midrule
        Comprehensiveness 
        & 44.8\%      & \textbf{55.2\%}     
        & 37.6\%      & \textbf{62.4\%}     
        & 18.8\%      & \textbf{81.2\%}
        & 32.0\%      & \textbf{68.0\%}
        & 28.4\%      & \textbf{71.6\%}\\
        Diversity 
        & 45.2\%      & \textbf{54.8\%}     
        & 49.2\%      & \textbf{50.8\%}     
        & 9.6\%       & \textbf{90.4\%}
        & 35.6\%      & \textbf{64.4\%}
        & 32.0\%      & \textbf{68.0\%}\\
        Empowerment 
        & 47.2\%      & \textbf{52.8\%}     
        & 41.2\%      & \textbf{58.8\%}     
        & 15.6\%      & \textbf{84.4\%}
        & 31.2\%      & \textbf{68.8\%}
        & 30.4\%      & \textbf{69.6\%}\\
        Overall 
        & 43.6\%      & \textbf{56.4\%}     
        & 40.0\%      & \textbf{60.0\%}     
        & 16.4\%      & \textbf{83.6\%}
        & 30.0\%      & \textbf{70.0\%}
        & 27.6\%      & \textbf{72.4\%}\\
        \cmidrule(lr){2-3} \cmidrule(lr){4-5} \cmidrule(lr){6-7} \cmidrule(lr){8-9} \cmidrule(lr){10-11} 
        & LightRAG & \textbf{FG-RAG}
        & LightRAG & \textbf{FG-RAG}
        & LightRAG & \textbf{FG-RAG}
        & LightRAG & \textbf{FG-RAG}
        & LightRAG & \textbf{FG-RAG} \\
        \midrule
        Comprehensiveness 
        & 46.4\%      & \textbf{53.6\%}     
        & 31.2\%      & \textbf{68.8\%}     
        & 40.4\%      & \textbf{59.6\%}
        & 41.2\%      & \textbf{58.8\%}
        & 29.6\%      & \textbf{70.4\%}\\
        Diversity 
        & 47.6\%      & \textbf{52.4\%}     
        & 46.8\%      & \textbf{53.2\%}     
        & 38.4\%      & \textbf{61.6\%}
        & \textbf{57.2\%}      & 42.8\%
        & 41.6\%      & \textbf{58.4\%}\\
        Empowerment 
        & 45.6\%      & \textbf{54.4\%}     
        & 30.8\%      & \textbf{69.2\%}     
        & 32.8\%      & \textbf{67.2\%}
        & 37.2\%      & \textbf{62.8\%}
        & 29.2\%      & \textbf{70.8\%}\\
        Overall 
        & 46.4\%      & \textbf{53.6\%}     
        & 32.0\%      & \textbf{68.0\%}     
        & 35.2\%      & \textbf{64.8\%}
        & 38.4\%      & \textbf{61.6\%}
        & 28.0\%      & \textbf{72.0\%}\\
        \cmidrule(lr){2-3} \cmidrule(lr){4-5} \cmidrule(lr){6-7} \cmidrule(lr){8-9} \cmidrule(lr){10-11} 
        & GraphRAG & \textbf{FG-RAG}
        & GraphRAG & \textbf{FG-RAG}
        & GraphRAG & \textbf{FG-RAG}
        & GraphRAG & \textbf{FG-RAG}
        & GraphRAG & \textbf{FG-RAG} \\
        \midrule
        Comprehensiveness 
        & 29.6\%      & \textbf{70.4\%}     
        & 30.8\%      & \textbf{69.2\%}     
        & 17.2\%      & \textbf{82.8\%}
        & 22.0\%      & \textbf{78.0\%}
        & 22.0\%      & \textbf{78.0\%}\\
        Diversity 
        & 40.8\%      & \textbf{59.2\%}     
        & 41.6\%      & \textbf{58.4\%}     
        & 25.6\%      & \textbf{74.4\%}
        & 29.2\%      & \textbf{70.8\%}
        & 27.6\%      & \textbf{72.4\%}\\
        Empowerment 
        & 18.8\%      & \textbf{81.2\%}     
        & 25.6\%      & \textbf{74.4\%}     
        & 13.2\%      & \textbf{86.8\%}
        & 16.8\%      & \textbf{83.2\%}
        & 18.4\%      & \textbf{81.6\%}\\
        Overall 
        & 24.8\%      & \textbf{75.2\%}     
        & 29.6\%      & \textbf{70.4\%}     
        & 15.2\%      & \textbf{84.8\%}
        & 19.6\%      & \textbf{80.4\%}
        & 20.8\%      & \textbf{79.2\%}\\
        \midrule
        \multirow{2}{*}{\textbf{qwen2.5-3b-instruct}} % 合并第一、二行的第一列，并填写内容
        & \multicolumn{2}{c}{\textbf{Agriculture}} 
        & \multicolumn{2}{c}{\textbf{Art}} 
        & \multicolumn{2}{c}{\textbf{CS}} 
        & \multicolumn{2}{c}{\textbf{Mix}}
        & \multicolumn{2}{c}{\textbf{News}} \\ 
        \cmidrule(lr){2-3} \cmidrule(lr){4-5} \cmidrule(lr){6-7} \cmidrule(lr){8-9} \cmidrule(lr){10-11} 
        & NaiveRAG & \textbf{FG-RAG}
        & NaiveRAG & \textbf{FG-RAG}
        & NaiveRAG & \textbf{FG-RAG}
        & NaiveRAG & \textbf{FG-RAG} 
        & NaiveRAG & \textbf{FG-RAG} \\
        \midrule
        Comprehensiveness 
        & 38.0\%      & \textbf{62.0\%}     
        & 28.8\%      & \textbf{71.2\%}     
        & 9.2\%       & \textbf{90.8\%}
        & 20.0\%      & \textbf{80.0\%}
        & 28.8\%      & \textbf{71.2\%}\\
        Diversity 
        & 34.4\%      & \textbf{65.6\%}     
        & 31.6\%      & \textbf{68.4\%}     
        & 3.2\%       & \textbf{96.8\%}
        & 18.8\%      & \textbf{81.2\%}
        & 19.2\%      & \textbf{80.8\%}\\
        Empowerment 
        & 34.0\%      & \textbf{66.0\%}     
        & 26.8\%      & \textbf{73.2\%}     
        & 6.4\%       & \textbf{93.6\%}
        & 16.8\%      & \textbf{83.2\%}
        & 24.4\%      & \textbf{75.6\%}\\
        Overall 
        & 34.8\%      & \textbf{65.2\%}     
        & 25.6\%      & \textbf{74.4\%}     
        & 6.4\%       & \textbf{93.6\%}
        & 16.4\%      & \textbf{83.6\%}
        & 25.6\%      & \textbf{74.4\%}\\
        \cmidrule(lr){2-3} \cmidrule(lr){4-5} \cmidrule(lr){6-7} \cmidrule(lr){8-9} \cmidrule(lr){10-11} 
        & LightRAG & \textbf{FG-RAG}
        & LightRAG & \textbf{FG-RAG}
        & LightRAG & \textbf{FG-RAG}
        & LightRAG & \textbf{FG-RAG}
        & LightRAG & \textbf{FG-RAG} \\
        \midrule
        Comprehensiveness 
        & 47.2\%      & \textbf{52.8\%}     
        & 38.4\%      & \textbf{61.6\%}     
        & 26.8\%      & \textbf{73.2\%}
        & 40.8\%      & \textbf{59.2\%}
        & 44.0\%      & \textbf{56.0\%}\\
        Diversity 
        & 42.4\%      & \textbf{57.6\%}     
        & 38.0\%      & \textbf{62.0\%}     
        & 21.6\%      & \textbf{78.4\%}
        & 40.0\%      & \textbf{60.0\%}
        & \textbf{50.4\%}      & 49.6\%\\
        Empowerment 
        & 44.4\%      & \textbf{55.6\%}     
        & 33.6\%      & \textbf{66.4\%}     
        & 21.2\%      & \textbf{78.8\%}
        & 32.4\%      & \textbf{67.6\%}
        & 40.8\%      & \textbf{59.2\%}\\
        Overall 
        & 46.0\%      & \textbf{54.0\%}     
        & 36.4\%      & \textbf{63.6\%}     
        & 23.6\%      & \textbf{76.4\%}
        & 35.6\%      & \textbf{64.4\%}
        & 43.2\%      & \textbf{56.8\%}\\
        \cmidrule(lr){2-3} \cmidrule(lr){4-5} \cmidrule(lr){6-7} \cmidrule(lr){8-9} \cmidrule(lr){10-11} 
        & GraphRAG & \textbf{FG-RAG}
        & GraphRAG & \textbf{FG-RAG}
        & GraphRAG & \textbf{FG-RAG}
        & GraphRAG & \textbf{FG-RAG}
        & GraphRAG & \textbf{FG-RAG} \\
        \midrule
        Comprehensiveness 
        & 28.0\%      & \textbf{72.0\%}     
        & 28.2\%      & \textbf{71.8\%}     
        & 23.6\%      & \textbf{76.4\%}
        & 14.4\%      & \textbf{85.6\%}
        & 22.0\%      & \textbf{78.0\%}\\
        Diversity 
        & 39.2\%      & \textbf{60.8\%}     
        & 33.6\%      & \textbf{66.4\%}     
        & 34.0\%      & \textbf{66.0\%}
        & 22.8\%      & \textbf{77.2\%}
        & 26.0\%      & \textbf{74.0\%}\\
        Empowerment 
        & 21.2\%      & \textbf{78.8\%}     
        & 16.8\%      & \textbf{83.2\%}     
        & 17.2\%      & \textbf{82.8\%}
        & 10.0\%      & \textbf{90.0\%}
        & 16.8\%      & \textbf{83.2\%}\\
        Overall 
        & 25.2\%      & \textbf{74.8\%}     
        & 20.4\%      & \textbf{79.6\%}     
        & 21.6\%      & \textbf{78.4\%}
        & 11.6\%      & \textbf{88.4\%}
        & 17.6\%      & \textbf{82.4\%}\\
        \midrule
        \multirow{2}{*}{\textbf{qwen2.5-1.5b-instruct}} % 合并第一、二行的第一列，并填写内容
        & \multicolumn{2}{c}{\textbf{Agriculture}} 
        & \multicolumn{2}{c}{\textbf{Art}} 
        & \multicolumn{2}{c}{\textbf{CS}} 
        & \multicolumn{2}{c}{\textbf{Mix}}
        & \multicolumn{2}{c}{\textbf{News}} \\ 
        \cmidrule(lr){2-3} \cmidrule(lr){4-5} \cmidrule(lr){6-7} \cmidrule(lr){8-9} \cmidrule(lr){10-11} 
        & NaiveRAG & \textbf{FG-RAG}
        & NaiveRAG & \textbf{FG-RAG}
        & NaiveRAG & \textbf{FG-RAG}
        & NaiveRAG & \textbf{FG-RAG} 
        & NaiveRAG & \textbf{FG-RAG} \\
        \midrule
        Comprehensiveness 
        & 44.0\%      & \textbf{56.0\%}     
        & 36.0\%      & \textbf{64.0\%}     
        & 35.2\%      & \textbf{64.8\%}
        & 44.4\%      & \textbf{55.6\%}
        & 31.6\%      & \textbf{68.4\%}\\
        Diversity 
        & 42.4\%      & \textbf{57.6\%}     
        & 39.6\%      & \textbf{60.4\%}     
        & 26.4\%      & \textbf{73.6\%}
        & 47.6\%      & \textbf{52.4\%}
        & 36.4\%      & \textbf{63.6\%}\\
        Empowerment 
        & 44.4\%      & \textbf{55.6\%}     
        & 36.8\%      & \textbf{63.2\%}     
        & 29.2\%      & \textbf{70.8\%}
        & 46.8\%      & \textbf{53.2\%}
        & 31.2\%      & \textbf{68.8\%}\\
        Overall 
        & 43.6\%      & \textbf{56.4\%}     
        & 34.4\%      & \textbf{65.6\%}     
        & 30.8\%      & \textbf{69.2\%}
        & 44.0\%      & \textbf{56.0\%}
        & 31.2\%      & \textbf{68.8\%}\\
        \cmidrule(lr){2-3} \cmidrule(lr){4-5} \cmidrule(lr){6-7} \cmidrule(lr){8-9} \cmidrule(lr){10-11} 
        & LightRAG & \textbf{FG-RAG}
        & LightRAG & \textbf{FG-RAG}
        & LightRAG & \textbf{FG-RAG}
        & LightRAG & \textbf{FG-RAG}
        & LightRAG & \textbf{FG-RAG} \\
        \midrule
        Comprehensiveness 
        & 42.4\%      & \textbf{57.6\%}     
        & 31.6\%      & \textbf{68.4\%}     
        & 31.6\%      & \textbf{68.4\%}
        & 38.4\%      & \textbf{61.6\%}
        & 32.4\%      & \textbf{67.6\%}\\
        Diversity 
        & 35.6\%      & \textbf{64.4\%}     
        & 35.2\%      & \textbf{64.8\%}     
        & 28.8\%      & \textbf{71.2\%}
        & 45.2\%      & \textbf{54.8\%}
        & 37.6\%      & \textbf{62.4\%}\\
        Empowerment 
        & 38.4\%      & \textbf{61.6\%}     
        & 32.0\%      & \textbf{68.0\%}     
        & 26.0\%      & \textbf{74.0\%}
        & 32.8\%      & \textbf{67.2\%}
        & 29.2\%      & \textbf{70.8\%}\\
        Overall 
        & 40.8\%      & \textbf{59.2\%}     
        & 32.4\%      & \textbf{67.6\%}     
        & 27.2\%      & \textbf{72.8\%}
        & 36.4\%      & \textbf{63.6\%}
        & 32.0\%      & \textbf{68.0\%}\\
        \cmidrule(lr){2-3} \cmidrule(lr){4-5} \cmidrule(lr){6-7} \cmidrule(lr){8-9} \cmidrule(lr){10-11} 
        & GraphRAG & \textbf{FG-RAG}
        & GraphRAG & \textbf{FG-RAG}
        & GraphRAG & \textbf{FG-RAG}
        & GraphRAG & \textbf{FG-RAG}
        & GraphRAG & \textbf{FG-RAG} \\
        \midrule
        Comprehensiveness 
        & 21.6\%      & \textbf{78.4\%}     
        & 20.8\%      & \textbf{79.2\%}     
        & 17.6\%      & \textbf{82.4\%}
        & 32.0\%      & \textbf{68.0\%}
        & 18.0\%      & \textbf{82.0\%}\\
        Diversity 
        & 36.0\%      & \textbf{64.0\%}     
        & 30.8\%      & \textbf{69.2\%}     
        & 35.2\%      & \textbf{64.8\%}
        & 47.6\%      & \textbf{52.4\%}
        & 26.0\%      & \textbf{74.0\%}\\
        Empowerment 
        & 19.2\%      & \textbf{80.8\%}     
        & 16.8\%      & \textbf{83.2\%}     
        & 16.4\%      & \textbf{83.6\%}
        & 28.4\%      & \textbf{71.6\%}
        & 16.0\%      & \textbf{84.0\%}\\
        Overall 
        & 21.6\%      & \textbf{78.4\%}     
        & 21.2\%      & \textbf{78.8\%}     
        & 17.6\%      & \textbf{82.4\%}
        & 33.6\%      & \textbf{66.4\%}
        & 16.0\%      & \textbf{84.0\%}\\
        \bottomrule
    \end{tabular}
    }
    \caption{Comparison of FG-RAG's win rate with other baselines across LLMs, metrics, and datasets.}
    \label{tab:detail}
\end{table*}

\clearpage
\newpage


\begin{table*}[!t]
\centering
\vspace{-0.05in}
\begin{tabular}{|c|p{14cm}|}
\hline
\textbf{Query} & How have artistic crossovers influenced political and social relations historically? \\ \hline
\textbf{FG-RAG} & ... The collaboration between diverse creative expressions has historically fostered dialogue around contemporary political climates and cultural dynamics. 
1. \textbf{Artistic Expression as a Catalyst for Change} [...]  
2. \textbf{Public Art and Community Engagement} [...]  
3. \textbf{The Role of Historical Figures} [...]  
4. \textbf{Gender and Cultural Representation} [...] 
\textbf{Conclusion}: In conclusion, the historical interplay between artistic crossovers and political/social relations ... \\ \hline
\textbf{NaiveRAG} &
... It embodies cultural responses to political realities and social engagements. 
1. \textbf{The Role of Modernism} [...]  
2. \textbf{The Shift to Postmodernism} [...]  
3. \textbf{Cultural Politics and Resistance} [...]  
\textbf{Conclusion}: Artistic crossovers have historically influenced political and social relations by enabling connections ... \\ \hline
\textbf{LLM Decision} & \textbf{Comprehensiveness}:  
``Winner": ``Answer 1 (FG-RAG)", % \newline

``Explanation": ``Answer 1 provides a detailed exploration of how artistic crossovers influence political and social relations, covering a variety of aspects  ..."

\textbf{Diversity}:  
``Winner": ``Answer 1 (FG-RAG)", %

``Explanation": ``While both answers engage with multiple artistic movements, Answer 1 presents a wider range of perspectives by including diverse artistic forms, historical figures, and gender-based contributions. It highlights  ..."

\textbf{Empowerment}:  
``Winner": ``Answer 1 (FG-RAG)", % \newline

``Explanation": ``Answer 1 effectively empowers the reader by demonstrating how art can instigate discussions and change in society. It explains artistic crossovers and their implications for ..."

\textbf{Overall Winner}:  
``Winner": ``Answer 1 (FG-RAG)",  %\newline

``Explanation": ``Answer 1 is the overall winner due to its depth of detail, breadth of perspectives, and strong emphasis on empowerment. It effectively articulates the historical significance ..." \\ \hline

\end{tabular}
\vspace{-0.05in}
\caption{A representative case comparing NaiveRAG to our FG-RAG method.}
\label{tab:naive}
\vspace{-0.05in}
\end{table*}

\begin{table*}[!t]
\centering
\vspace{-0.05in}
\begin{tabular}{|c|p{14cm}|}
\hline
\textbf{Query} & How have artistic crossovers influenced political and social relations historically? \\ \hline
\textbf{FG-RAG} & ... The collaboration between diverse creative expressions has historically fostered dialogue around contemporary political climates and cultural dynamics. 
1. \textbf{Artistic Expression as a Catalyst for Change} [...]  
2. \textbf{Public Art and Community Engagement} [...]  
3. \textbf{The Role of Historical Figures} [...]  
4. \textbf{Gender and Cultural Representation} [...] 
\textbf{Conclusion}: In conclusion, the historical interplay between artistic crossovers and political/social relations ... \\ \hline
\textbf{GraphRAG} &
... These intersections can be observed across various historical contexts, illustrating how art has been utilized to advocate for social change, influence public perception, and foster diplomatic relations. 
1. \textbf{Art as a Catalyst for Social Change} [...]  
2. \textbf{Political Commentary Through Artistic Expression} [...]  
3. \textbf{Diplomatic Relations and Cultural Exchange} [...]  
\textbf{Conclusion}: In summary, artistic crossovers have historically influenced political and social relations by serving as vehicles for social change, platforms for ... \\ \hline
\textbf{LLM Decision} & \textbf{Comprehensiveness}:  
``Winner": ``Answer 1 (FG-RAG)", % \newline

``Explanation": ``Answer 1 provides a detailed exploration of various aspects of how artistic crossovers influence political and social relations. It discusses specific artists and movements, such as Ibsen and Rubens, while also addressing ..."

\textbf{Diversity}:  
``Winner": ``Answer 1 (FG-RAG)", %

``Explanation": ``Answer 1 showcases a wider range of artistic forms and historical examples, including theatre, public art, and unique contributions from women artists. In contrast, Answer 2 focuses more heavily on ..."

\textbf{Empowerment}:  
``Winner": ``Answer 1 (FG-RAG)", % \newline

``Explanation": ``Answer 1 effectively helps the reader understand the significant role of artistic expressions in societal and political contexts, encouraging them to reflect on the lasting impact of these artworks. It emphasizes ..."

\textbf{Overall Winner}:  
``Winner": ``Answer 1 (FG-RAG)",  %\newline

``Explanation": ``Overall, Answer 1 emerges as the stronger response due to its comprehensive detail, diverse range of examples and perspectives, and its ability to empower ..." \\ \hline

\end{tabular}
\vspace{-0.05in}
\caption{A representative case comparing GraphRAG to our FG-RAG method.}
\label{tab:graph}
\vspace{-0.05in}
\end{table*}

\begin{table*}[!t]
\centering
\vspace{-0.05in}
\begin{tabular}{|c|p{14cm}|}
\hline
\textbf{Query} & How have artistic crossovers influenced political and social relations historically? \\ \hline
\textbf{FG-RAG} & ... The collaboration between diverse creative expressions has historically fostered dialogue around contemporary political climates and cultural dynamics. 
1. \textbf{Artistic Expression as a Catalyst for Change} [...]  
2. \textbf{Public Art and Community Engagement} [...]  
3. \textbf{The Role of Historical Figures} [...]  
4. \textbf{Gender and Cultural Representation} [...] 
\textbf{Conclusion}: In conclusion, the historical interplay between artistic crossovers and political/social relations ... \\ \hline
\textbf{FastGraphRAG} &
... This phenomenon can be traced across different eras and movements, where art not only served as a medium of expression but also as a catalyst for change and reflection of societal dynamics. 
1. \textbf{Art as a Reflection of Political Change} [...] 
2. \textbf{Influence of the Harlem Renaissance} [...]  
3. \textbf{The Role of Baroque and Renaissance Art} [...] 
4. \textbf{Postmodernism and Cultural Critique} [...] 
\textbf{Conclusion}: In conclusion, the interplay between artistic crossover and political as well as social relations throughout history demonstrates the profound ... \\ \hline
\textbf{LLM Decision} & \textbf{Comprehensiveness}:  
``Winner": ``Answer 1 (FG-RAG)", % \newline

``Explanation": ``Answer 1 provides a detailed exploration of various aspects of how artistic crossovers influence political and social relations. It discusses specific artists and movements, such as Ibsen and Rubens, while also addressing ..."

\textbf{Diversity}:  
``Winner": ``Answer 1 (FG-RAG)", %

``Explanation": ``Answer 1 showcases a richer diversity of perspectives by incorporating a wider range of artistic examples, historical contexts, and gender considerations. It discusses the role ..."

\textbf{Empowerment}:  
``Winner": ``Answer 1 (FG-RAG)", % \newline

``Explanation": ``Answer 1 empowers readers by offering them a deeper understanding of how art can reflect societal issues and incite change. It encourages critical thinking about the role ..."

\textbf{Overall Winner}:  
``Winner": ``Answer 1 (FG-RAG)",  %\newline

``Explanation": ``Answer 1 is the overall winner as it excels in all three criteria—comprehensiveness, diversity, and empowerment. It provides a thorough and nuanced exploration of the intersections between ..." \\ \hline

\end{tabular}
\vspace{-0.05in}
\caption{A representative case comparing FastGraphRAG to our FG-RAG method.}
\label{tab:fastgraph}
\vspace{-0.05in}
\end{table*}

\end{document}